\documentclass{knawproc}
\usepackage{epsfig}

\def\,{\kern .16667em}

\begin{document}

\begin{opening}

\title{Large Scale Structure: \\
Setting the Stage for the Galaxy Formation Saga}

\author{Rien van de Weygaert}
\addresses{%
  Kapteyn Institute, University of Groningen, P.O. Box 800, 9700 AV Groningen, 
The Netherlands\\}
\runningtitle{Large Scale Structure: the stage for galaxy formation}
\runningauthor{van de Weygaert}

\end{opening}


\begin{abstract}
Over the past three decades the established view of a nearly homogeneuous, 
featureless Universe on scales larger than a few Megaparsec has been 
completely overhauled. In particular through the advent of ever larger 
galaxy redshift surveys we were revealed a galaxy distribution displaying an 
intriguing cellular pattern in which filamentary and wall-like structures, 
as well as huge regions devoid of galaxies, are amongst the most 
conspicuous morphological elements. 

In this contribution we will provide an overview of the present observational 
state of affairs concerning the distribution of galaxies and the structure 
traced out by the matter distribution in our Universe. In conjunction with 
the insight on the dynamics of the structure formation process obtained 
through the mapping of the peculiar velocities of galaxies 
in our local Universe and the information on the embryonic 
circumstances that prevailed at the epoch of Recombination yielded by the 
various Cosmic Microwave Background experiments, we seek to arrive at 
a more or less compelling theoretical framework of structure formation.
The main aspects of this framework of the rise of structure through 
gravitational instability can probably be most readily appreciated through 
illustrative examples of various scenarios, as for instance provided by 
some current state-of-the-art N-body simulations. 

We will subsequently wrap up the observational and theoretical evidence 
for the emergence and evolution of structure in the Universe by sketching 
the stage for 
the ultimate Holy Grail of late 20th century astrophysics, understanding 
the saga of the formation of what arguably are the most prominent and at 
the same time intoxicatingly beautiful and intriguing denizens of our 
Cosmos, the {\it galaxies}. 
\end{abstract}


\begin{figure}
\centering
\epsfysize=19.cm
\epsfbox{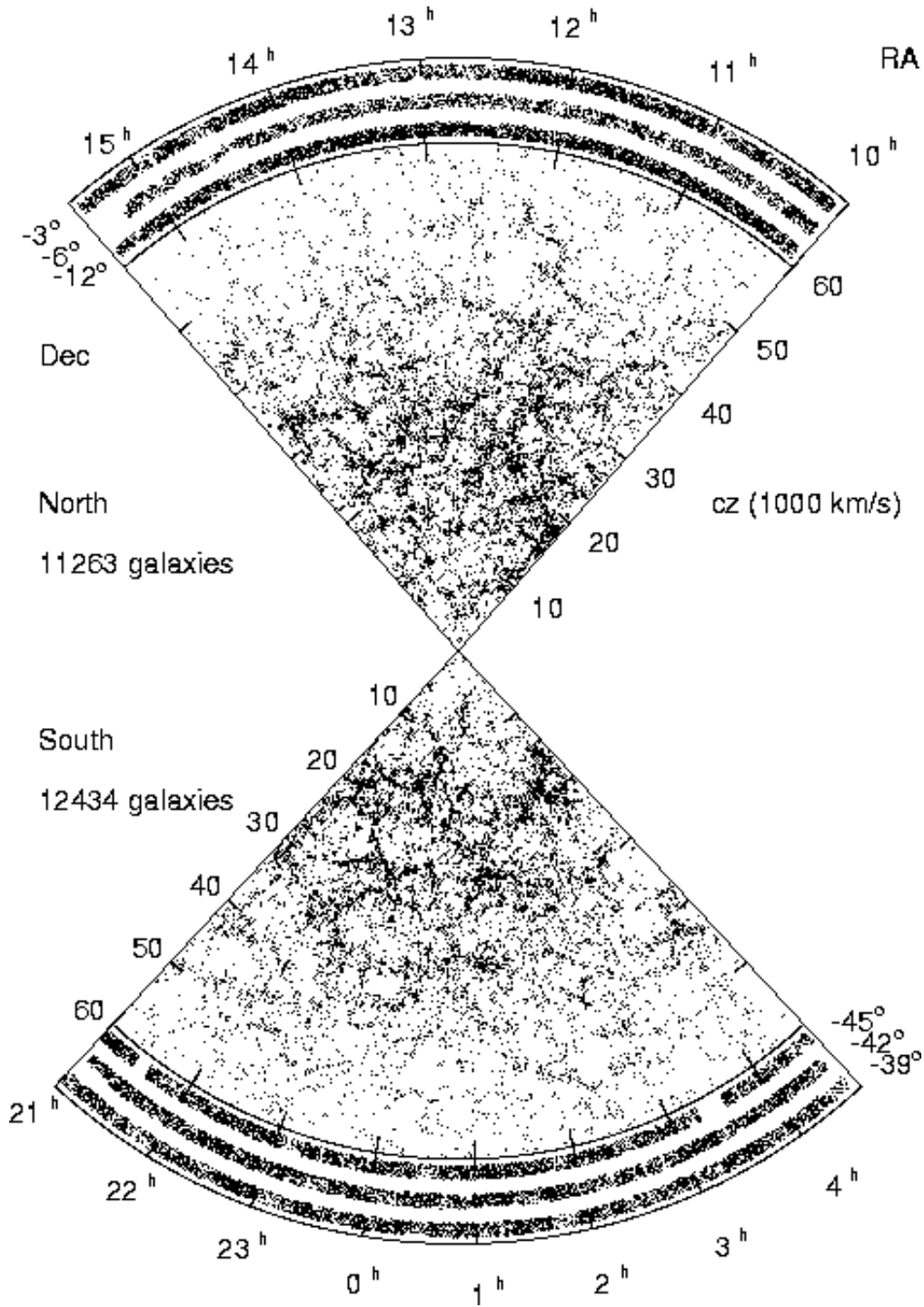}
\caption{The Las Campanas redshift survey. The figure displays 
the threedimensional position of 26418 galaxies in 6 thin strips on 
the sky. The total survey comprises nearly 700 square degrees, with 
each strip measuring a 1.5$^{\circ}$ $\times$ 80$^{\circ}$ region on 
the sky, with the survey extending out to an effective depth of 
approximately 300 Megaparsec. Clearly visilbe is the spongelike 
features into which the galaxies have organized themselves, with 
filaments and walls surrounding void regions with characteristic 
sizes in the order of 50 Mpc. Courtesy: Shectman, S., Schechter, P., 
Oemler, G., Kirshner, B., Tucker, D., Landy, S., Hashimoto, Y. \& Lin, H.}
\end{figure}

\section{Cosmic Minutiae: the origin of cosmic structure}
With the twentieth century drawing to a close we may be melancholic and 
see that human irrationality and evil ran havoc on an unprecedented 
scale, exposing the world to almost inconceivable levels of destruction, 
terror and suffering. On the other hand, we 
are equally justified in priding ourselves of participating in an era 
of unparalleled triumph of the human ratio, a century full of 
tremendous scientific progress and enlightement. Forever our century will 
figure as the one in which the human mind finally succeeded in unlocking  
the seals holding the secret to our cosmic origin. Cosmology finally broke 
away from its mythological roots, and for the first time since the Dawn 
of Civilization the centuries old quest for the origin of the world 
seems to have come across a well-founded and consistent answer, inscribed 
in the scientific epos of the ``Hot Big Bang'' theory. The success 
of these relativistic, homogeneous and isotropic Friedmann-Roberstson-Walker 
Universe models --- including the repercussions for the subsequent unfolding 
of the physical state of Universe --- in describing the structure and 
evolution properties of the Universe on global scales of many hundreds of Megaparsec and beyond, is truely enchanting. 

However, the Friedmann-Robertson-Walker Universe only represents the cosmos 
in its most global and universal context, it does not contain any 
explanation of its own origin, nor of the state or even the existence 
of its constituents. It does not go to any extent in explaining why 
the Universe is one harbouring a wealthy internal structure. Moreover, the 
past three decades have 
seen a radical revision of our view of the structural organization of matter 
in our Universe. The canonic view of a nearly homogeneuous, featureless 
Universe on scales larger than a few Megaparsec got completely overhauled 
into one displaying a baffling richness, populated by an astonishing 
variety of objects. As a matter of illustration, when turning to Figure 1, 
a map of the distribution of galaxies yielded by the Las Campanas survey 
(e.g. Landy et al. 1996),  
we see that the image of a homogeneous Universe is far besides truth, and 
cannot be anything but an approximation of reality, valid only for the 
Universe on global scales. Within cosmology the issue of the formation 
of structure has therefore gradually manoeuvred itself to the forefront of 
scientific interest. Not only because it fills in an obvious hiatus in the 
Friedmann-Robertson-Walker models, but also because we have come to 
realize that an understanding of the structure formation process is of 
key significance in unravelling the primordial physical processes 
determining the evolution and fate of the Universe itself. While 
we may have the impression that objects and structures ranging from 
planets, stars up to superclusters are but cosmic minutiae, we 
should realize that it is often minutiae that are the punctuation 
marks enabling us to systematize the flood of information reaching us 
from cosmic realms into new levels of insight. 

The central unsolved riddle of cosmology has therefore become the 
question of how the near perfectly homogeneous, featureless, extremely hot 
and dense early Universe gave rise to the wealth and variety of structure 
which make our cosmos into such a fascinating world to live in. 
Instrumental in solving this puzzle is the realization that our 
Universe still contains cosmological fossils, structures and physical 
properties that still contain 
traces of the processes that have been responsible for the emergence 
of all the objects and structures populating our cosmos.
The way in which matter has arranged itself on scales of a few 
up to several hundreds Megaparsec has evolved sufficiently far to yield 
observable manifestations of the growth process while its matter contents  
and internal motions have not yet been blended to such an extent that 
they do no longer contain any directly and objectively retrievable 
information on the structure formation processes. 

These fossils revealed themselves as cosmological research in the 
second half of the twentieth century came across the existence of 
a rich and beforehand unexpected organization of matter into structures 
over a large range of scales. The structure of these matter arrangements has 
been unveiled through the mapping of the distribution of the galaxy distribution, currently even over a substantial range of cosmic history. Moreover, 
by virtue of its imprint on passing radiation from objects in the background, 
snatching its Ly $\alpha$ photons, culminating into a forest of 
``Ly $\alpha$'' 
lines, we are even obtaining a reasonable idea of the distribution and 
physical state of diffuse matter populating the regions in between 
identifiable ``objects''. Even the embryonic state of the cosmic matter 
distribution has been exposed to scientific 
exploration through the detection of temperature fluctuations in 
the cosmic background radiation, following the ground-breaking efforts 
of the COBE satellite. We even have been able to obtain insight into 
the dynamics underlying the structure formation process, as meticulous 
and careful measurements of peculiar motions of galaxies yielded 
information on the forces that have been shaping the organization 
of matter. Moreover, stakes are high that exploitation of the gravitational 
impact on the path of photons, usually phrased as gravitational lensing, will 
enable us to get an unbiased view of the gravitational potential throughout theUniverse. 

While we are witnessing a continuously expanding inflow of new data on 
cosmic structure, a sensible interpretation of these data can only 
be achieved through providing a general physical framework for their 
formation and evolution. Although there have been several theories 
around, over the past decade one has clearly obtained the lead, most 
data at least partially corroborating its implications, the theory 
of {\it Gravitational Instability}. 

\section{A Cosmological Footnote: Creation through Gravitational Instability}
The finding of COBE of very small fluctuations in the temperature of the 
microwave background radiation, and its interpretation in terms of 
slight variations of the gravitational potential at the surface 
of last scattering, is a remarkable confirmation of the general
theoretical framework of ``gravitational instability'' for cosmic
structure formation. According
to this theory the early universe was almost perfectly smooth except for 
tiny density variations with respect to the general background 
density of the universe and related tiny velocity perturbations 
with respect to the general Hubble expansion. Because slight density 
enhancements exert a slightly stronger gravitational attraction 
on the surrounding matter, they start to accrete material from 
its surroundings as long as pressure forces are not sufficient 
to counteract this infall. In this way the overdensity becomes even 
more overdense, and their gravitational influence even stronger. The 
denser it becomes the more it will accrete, resulting in an
instability which can ultimately cause the collapse of a density 
fluctuation to a gravitationally bound object. This generic process 
of structure formation is illustrated in Figure 2, displaying both 
a density fluctuation field, the corresponding force field and the 
subsequent displacement of parcels of matter. The size and mass of 
the object is of course dependent on the scale of the fluctuation. 
For example, galaxies are thought to have formed out of 
fluctuations on a scale of $\approx 0.5h^{-1}\hbox{Mpc}$, while 
clusters of galaxies have emerged out of fluctuations on a 
larger scale of $\approx 4h^{-1}\hbox{Mpc}$. The formation of 
voids fits in the same general scheme, having grown out of 
primordial underdensities in the matter distribution. 

Providing the general framework, the gravitational instability theory 
needs lots of details to be filled in before it can be considered a 
complete theory. There is of course the issue of the amount of 
matter represented by a density fluctuation, as more 
massive fluctuations will collapse sooner. Given the amplitude of 
the fluctuations, their total mass is determined by the average 
cosmological density, paramerized by $\Omega$. The very low 
value of the amplitude of the primordial density fluctuations 
inferred from the COBE MWB measurements is a strong argument in
favour of a high overall density of the universe. Otherwise, density 
fluctuations would simply not have had sufficient time to collapse 
on all the scales that nowadays are observed to exhibit so much 
structure. Also some other observational indications support 
a high value of $\Omega$, which has the important implication 
that most likely the major share of matter in the universe does not
consist of familiar baryons and leptons but of one or 
more as yet unidentified species of ``dark matter''. 

The nature and amount of dark matter is also of substantial influence
in determining the character of the initial density and fluctuation field, 
probably the most crucial issue in the structure formation saga. Rather 
than consisting of some isolated, well-defined and smooth density
peaks and dips, each of its own particular scale, the density field
can be thought of as a random superposition of fluctuations of various  
scales. It will therefore bear the character of a noise field, 
``a random field'', a random superposition of waves much like the
surface of the sea at rough weather. Evidently, the waves with the 
largest amplitude will collapse first. The character of the
density field evolution will then depend on the relative
amplitudes of the different waves. One extreme case is that of
small scale waves having by far the highest amplitude. 
They will collapse into virialized objects well before a larger scale 
perturbation, in which they are possibly embedded, starts to
collapse. Consequently, we will see a hierarchical or ``bottom-up''
build-up of structure, where small objects that formed first merge
into larger structures, which themselves merge to form galaxies,
cluster of galaxies, and so on. The other extreme is that of the 
case in which there are only perturbations on large scales, with no 
contributions from smaller scales. In such a ``top-down'' scenario 
the first emerging structures form through the collapse of those 
large scale perturbations. In the most popular versions of 
``top-down'' theories these objects would correspond to superclusters. 
Subsequently, smaller objects like galaxies have to form 
through the fragmentation of these collapsed large objects into 
smaller pieces, an as yet mostly ununderstood process in which 
non-gravitational gas processes play a key role. 

\begin{figure}
\centering
\epsfysize=15.5cm
\epsfbox{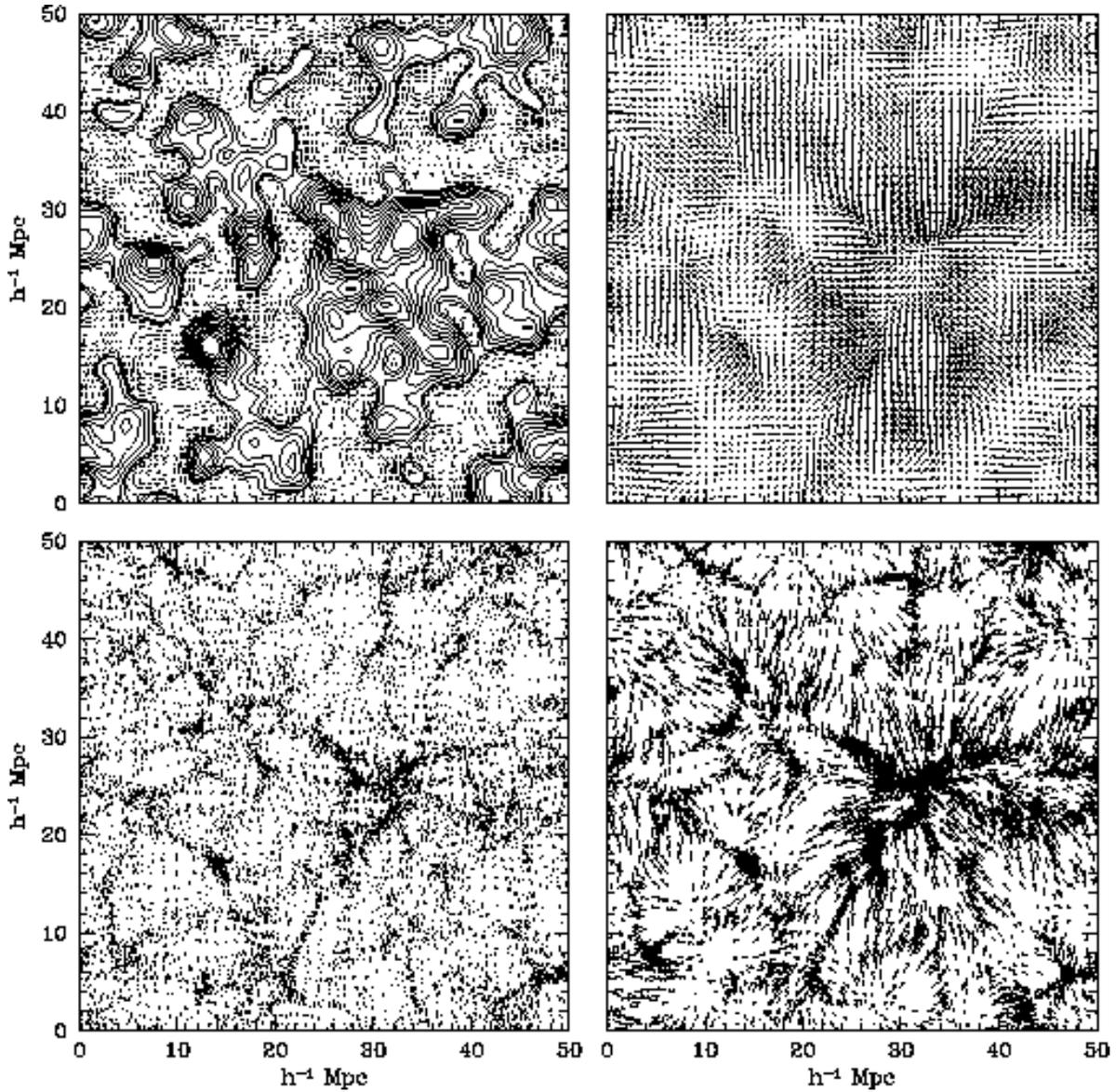}
\caption{Illustration of structure formation through gravitational 
instability. A cut through a random density field realization is 
displayed in the upper lefthand frame. The corresponding force 
field is shown in the upper righthand frame. This force field 
induces matter displacement, leading up to a distribution of matter 
shown in the lower lefthand corner. Compare this with the corresponding 
streaming velocities in the lower righthand frame.}
\end{figure}

The formation of anisotropic structural patterns in these random
density fields is the consequence of an additional characteristic property of 
gravitational collapse. Overdensities, on any scale and in any 
scenario, always collapse such that they become increasingly
anisotropic. At first they turn into a flattened ``pancake'', later 
possibly followed by contraction into an elongated filament or 
by full collapse into a virialized clump like a galaxy or a cluster. 
This tendency to collapse anisotropically is caused by the intrinsic 
primordial flattening of the overdensity as well as by the anisotropy
of the gravitational force field induced by the external matter 
distribution, i.e. by tidal forces. In the case of a pure hierarchical
scenario the amplitude of large scale overdensities will be so low
that they will not really have started their anisotropic collapse 
before the small scale overdensities have turned into high-density 
virialized clumps. Instead of appearing like a large coherent 
anisotropic structure the resulting large scale matter distribution 
will therefore more resemble a mere incoherent and shapeless density 
enhancement in the number of small clumps. On the other hand, in less 
extreme hierarchical scenarios large scale density fluctuations will
have an amplitude high enough such that by the time small scale clumps
have completely collapsed the large scale structure in which they are 
embedded will already have contracted substantially. In those cases we 
expect to see more or less coherent walls and filaments in which 
the small scale clumps stand out like beads on a string. Finally, in 
the most extreme ``top-down'' case we will only see the anisotropic
contraction of a large scale object like a supercluster. The resulting
pattern will be one of a network of filaments and walls without any 
internal structure. 

\section{Cosmic Symbiotics: Large Scale Structure and Galaxies}
From the preceding it has already been clear that galaxies play a 
central role in the efforts to map the structure of the Universe and 
to come to an encompassing theory of structure formation in the 
Universe. This immediately exposes a precarious issue within the 
whole framework of structure formation. What is the role of galaxies ? 
What is their nature ? 

Although obviously a subjective view, there is some right in 
considering galaxies are amongst the most beautiful and mesmerizing 
objects in the Universe. To some 
extent autarctic entities, cosmic cities harbouring and organizing all 
the ingredients necessary for bringing forth highly complex states of 
matter organization, like stars, planets, and even something we describe 
as ``life'', they are at the same time the beacons of the Universe. Mainly 
through their existence have we been able to study the structure of the 
Universe. 

Obviously, the hope is that by mapping the galaxy distribution we at the 
same time obtain a representative map of the matter distribution. However, 
this is only an assumption, a crucial one that theoretically has still not 
been justified. There are some a posteriori 
indications that it is indeed true on scales of a few Megaparsec and 
larger. However, no compelling theory exists of how and where galaxies 
would form within the large scale organization of matter. In order to 
extract firm conclusions in our search for cosmic structure formation, 
we therefore need to get a better understanding of the biased view 
that the distribution of galaxies represents, and hence of the 
process of galaxy formation. While this obviously is of prime importance 
in relating the galaxy and matter distribution, it is even true for 
the more objective, but less detailed, probe of the matter distribution 
offered by measured peculiar velocities. While we are probably not far 
from reality assuming that galaxies float along with all other matter 
currents in the Universe and therefore that their peculiar velocities 
represent excellent probes of the underlying velocity field, there still 
remains the possibility there is some level of ``velocity bias''. 

Here nature plays a trick on us. The formation of galaxies is 
not a purely gravitational matter, and therefore not one producing 
configurations readily retraceable to its cosmic origin and shaping 
agents. On the contrary, it is a highly 
complex and dissipative business consisting of a subtle interplay between 
gravitational, radiative and hydrodynamic processes on a range of 
scales. This complex interaction incorporates cooling processes of gas, 
ultimately leading to the formation of stars, feedback processes of exploding 
stars, enriching gas by heavier elements, while radiation emitted by stars and 
galactic and cosmic background will counter the cooling of gas. 
Hence rendering it impractable to try to understand galaxy formation 
on the basis of the structure and kinematics of galaxies themselves alone, 
the hope is that inferences about larger structures over a range of scales 
may be extrapolated to galaxy scale, providing the initial setting of 
protogalaxies. 

While this is arguably one of the most important yields of large scale 
structure studies, we are at the same time caught in a web as we have 
already seen that a complete and objective assessment makes it necessary 
to understand the process of galaxy formation. Big strides towards the 
ultimate resolution of the structure formation riddle therefore implies 
a hand-in-hand progress in theoretical understanding and observational 
indications. 

In an effort to paint the cosmic environment in which galaxies 
are assembled, in the hope of clearing up the environmental impact 
on the emergence of galaxies, we will first provide a description 
of the observed large scale patterns in the galaxy distribution.  
This will be followed by a short discussion of the efforts towards explaining 
the formation of these patterns through gravitational instability, 
utilizing ever more intricate simulations of the evolution of 
representative distributions of particles. In this way we hope to 
offer a framework for the incorporation and interpretation of the 
role of the extreme representatives of the galaxy population within 
the structure formation saga, so that we may have a better understanding of 
how to utilize the subjects of this meeting, radio galaxies at 
high redshift, as probes of their large scale environment.

\section{Foaming Delights: patterns in the cosmic matter distribution}
During the past three decades it were in particular major advances in 
telescope and detector technology that instigated a continuously stronger  
effort towards surveying and mapping the matter and galaxy distribution 
in the Universe. Penetrating previously unexplored swathes of the local 
Cosmos, systematic galaxy redshift surveys have uncovered the existence 
of an hitherto unexpected richly patterned and fascinating organization of 
matter on scales ranging from a few to even several hundred Megaparsec. 
Early hints (e.g. de Lapparent etal. 1986) for the existence of a foamlike 
textured galaxy distribution got strongly corroborated as larger 
and more ambitious surveys expanded their reach, establishing the 
image of a vast cosmic foamlike network ostensibly pervading nearly 
all of the visible Universe (see Fig. 1, Landy et al. 1996). 

\begin{figure}
\centering
\epsfysize=10.cm
\epsfbox{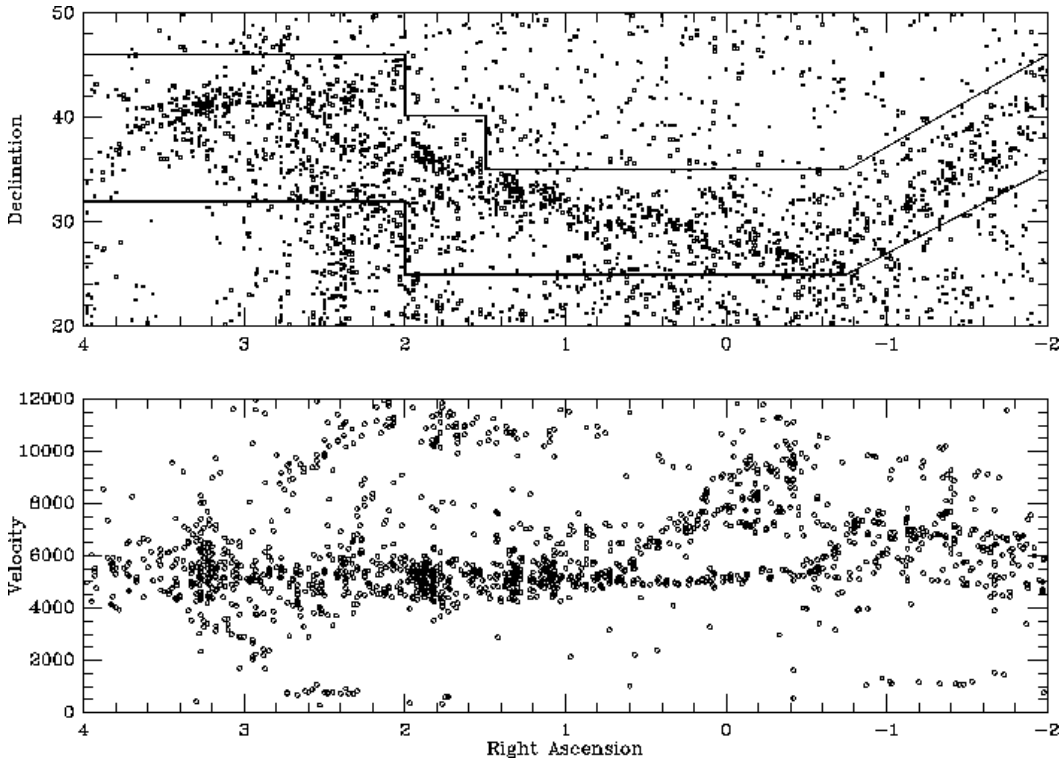}
\caption{The Perseus-Pisces supercluster chain of galaxies. Separate 
two-dimensional views of the galaxy distribution in the northern 
region of the Pisces-Perseus region. The upper panel shows the sky 
distribution of all galaxies in the overall northern survey sample 
of Wegner, Haynes \& Giovanelli (1993). The region believed to contain
the Pisces-Perseus main ridge is outlined. The lower panel shows the
two dimensional redshift distribution (right ascension-recession 
velocity $V_0$) for galaxies in the ridge region highlighted in the 
upper panel. From Giovanelli \& Haynes 1996, kindly provided by 
M. Haynes.} 
\end{figure}

The frothy geometry is evidently one of the most prominent aspects of 
the cosmic fabric, highlighted by galaxies populating 
huge {\it filamentary} and 
{\it wall-like} structures, the sizes of the most conspicuous one 
regularly exceeding 100h$^{-1}$ Mpc. The closest and best studied of these 
massive anisotropic matter concentrations can be identified 
with known supercluster complexes, enormous structures comprising one or more 
rich clusters of galaxies and a plethora of more modestly sized clumps 
of galaxies. Both our Local Group and the Virgo cluster are members of 
such a structure, the Local Supercluster, a huge flattened 
concentration of about fifty groups of galaxies in which the Virgo 
cluster is the dominant and central agglomeration. The Local supercluster 
is but a modest specimen of its class, dominated by only one rich 
cluster. A far more prominent example of a supercluster, and arguably 
more canonic in terms of morphological character, is the Perseus-Pisces 
supercluster (see Fig. 3). Due to its relative closeness (approximately 
55$h^{-1}$ Mpc), its characteristic and salient filamentary geometry, and 
its favourable orientation perpendicular to the line of sight, it has 
become one of the best mapped and meticulosly studied superclusters. It is 
a huge conglomeration of galaxies that clearly stands out on the sky. The 
boundary of the supercluster on the northern side is formed by the filament 
running southwestward from the Perseus cluster, a majestic chain of galaxies 
of truly impressive proportions. It has a length of at least 50$h^{-1}$ Mpc 
and a width 
of about 5$h^{-1}$Mpc. The ridge possibly extends even further out to a total 
length of 140$h^{-1}$ Mpc, although obscuration by the Galactic Disk prevents 
firm conclusions on this point. Along the major ridge we see a more or less 
continuous arrangement of high density clusters and groups, of which the most 
notable ones are the Perseus cluster itself (Abell 462), Abell 347 and 
Abell 262. An exquisit impression of its structure can be obtained from 
the 21 cm line redshift survey of some 5000 late-type galaxies in the Perseus 
region, by Giovanelli, Haynes and collaborators (see e.g. Wegner, Haynes 
and Giovanelli, 1993, and Fig. 3), the most detailed study of the region 
currently available. In addition to the presence of such huge filaments 
we can also discern vast planar assemblies in the galaxy distribution. A 
striking example of its kind is the {\it Great Wall} which was 
identified through the CfA2 survey (Geller \& Huchra 1989). It constitutes 
a huge planar assembly of galaxies with dimensions that are estimated to be 
of the order of $60h^{-1} \times 170h^{-1} \times 5h^{-1}$ Mpc, 
which has the Coma cluster of galaxies as its most prominent density 
enhancement. Another huge wall of galaxies in our cosmic neighbourhood 
has been found on the southern hemisphere (e.g. Da Costa 1993), adding 
to the impression of them being ubiquitous elements of cosmic 
structure. This impression got even more convincing support after the 
publication of the results of the deeper Las Campanas redshift survey 
(Fig. 1). Its chart of 26,000 galaxy locations in six thin strips on 
the sky, extending out to a redshift of $z \sim 0.1$, currently represents the 
best and most representative impression of cosmic structure available. In 
the near future we can look forward to considerable extensions of the 
cosmic atlas. The 2dF and Sloan redshift surveys have embarked on a majestic 
enterprise to probe the galaxy distribution of the Universe in hitherto 
unexplored regions of cosmic territory, out to scales of 
$\sim 1000h^{-1}\rm{Mpc}$ (see e.g. Lahav 1995, and 
website http://msowww.anu.edu.au/~colless/2dF/ for further details and 
even some recent results of the 2dF survey, and Gunn \& Weinberg 1995, 
Margon 1998 and website http://www-sdss.fnal.gov:8000/ for details and 
updates of the Sloan SDSS redshift survey). The compilation of more than 
a million galaxies they strive after will for the first time produce 
truely uniform and representative samples of our cosmic 
environment, a true voyage of discovery ... 

Not only do we come across filamentary and planar mass concentrations. In 
fact, perhaps one of the most intriguing discoveries emanating from 
extensive redshift surveys has been the existence of large {\it voids} 
in the galaxy distribution, enormous regions, sometimes up to tens of 
Megaparsec in extent, wherein few or no galaxies are found. The Bo\"otes 
voids in the KOSS redshift surveys (Kirshner et al. 1981, 1987) was the first 
of its kind to attract the attention. It is an almost completely 
empty spherical region (however, see Szomoru 1995) with a diameter of 
around $60h^{-1}$Mpc and is still regarded as the canonical example. Various 
redshift surveys covering large parts of the local Universe have 
shown that voids with sizes typically in the range of $20-50h^{-1}$ Mpc 
are a common feature in the galaxy distribution, at least up to a 
redshift of $z \sim 0.5$ (e.g. see Vogeley, Geller \& Huchra 1991 and 
Bellanger \& De Lapparent 1995). 

We may therefore conclude that filaments, walls and voids are eminent 
structural elements of the galaxy distribution. Moreover, 
a careful assessment of their distribution throughout space also shows them 
not to be merely independently and randomly scattered objects. On the 
contrary, the galaxy maps clearly reveal the voids to be 
generically associated with surrounding density enhancements. In other 
words, the voids, filaments and walls are not only outstanding components 
of the galaxy distribution. They also conspire by weaving themselves into the 
beautiful {\it foamlike} tapestry that permeates our universe wherever we 
turn our gaze (e.g. Fig. 1). Within the framework traced out by the galaxy 
distribution they are both contrasting as well as complementary ingredients, 
with the vast under-populated regions, (the {\it voids}), being surrounded by 
{\it walls} and {\it filaments}. At the intersections of the latter we often 
find the most prominent density enhancements in our universe, the 
{\it clusters} of galaxies. 

Within the scheme of galaxy clustering and structure formation these dense and 
rich clusters stand out as the apogee of objects 
that can still be considered individually distinguishable entities. 
Being the dwelling sites of sometimes up to thousands of galaxies, they  
constitute the most massive collapsed and virialized matter condensations
in the Universe. They appear to populate the high-mass tail of a wide 
spectrum of galaxy assemblies, from small groups of a few galaxies, 
via somewhat more substantial groups like our own Local Group up to 
the true giants like the Virgo Cluster or the even more majestic Coma 
Cluster. The majority of these groups and clusters are strewn over the 
foamlike network of filaments and walls, constituting the occasional 
density enhancements rendering these structures their often irregular 
appearance. Generically, these groups are therefore seen to concentrate 
along the high density ridges of the cosmic network, leading up to the 
sites where several filaments and walls intersect, often highlighted by 
the presence of one or more massive clusters. Moreover, the fact that 
the groups and clusters seem to display a more pronounced concentration 
towards the walls and filaments of the cosmic foam than the galaxies 
themselves do is reflected quantitatively in the higher amplitude of 
their two-point correlation function. In other words, they seem to 
represent a more biased tracer of the underlying distribution of mass. 

The ubiquity of the characteristic frothy cosmic structure over vast expanses 
of the visible Universe has already been confirmed by the results of redshift 
surveys in very small regions of the sky out to huge depths, the 
``pencil beam redshift surveys'' in some cases 
probing to redshifts in the order of $z \sim 0.5$. Most famed amongst 
its peers is the pencil beam redshift survey by Broadhurst et al. (1990), 
whose conspicuous spiky redshift distribution along a direction towards the 
North Galactic and South Galactic pole at the time got an 
ambivalent reception of surprise mixed with scepsis. Nonetheless, later work 
only strenghtened the impression of huge peaks in the one-dimensional 
redshift probes. Comparison with shallower wide-angle surveys made 
clear that the identification of these redshift spikes with Great Walls 
such as the one revealed by the CfA2 survey was fully warranted, 
the spikes coinciding with the locations where the narrow redshift probes 
were piercing through the cosmic walls. Moreover, such deep 
pencil beam probes make clear that pronounced structures on scales 
in the order of 100 Mpc already existed at surprisingly early cosmic 
epochs and therefore argue for a surprisingly early development of 
structure organization in the Universe. The most astonishing recent 
corroboration for such early action is the recent statistical evidence 
(Steidel et al. 1998) for a substantial level of clustering in the 
population of the socalled Lyman break galaxies, at redshifts of 
even $z \sim 3$. 

\begin{figure}
\centering
\epsfxsize=15.5cm
\epsfbox{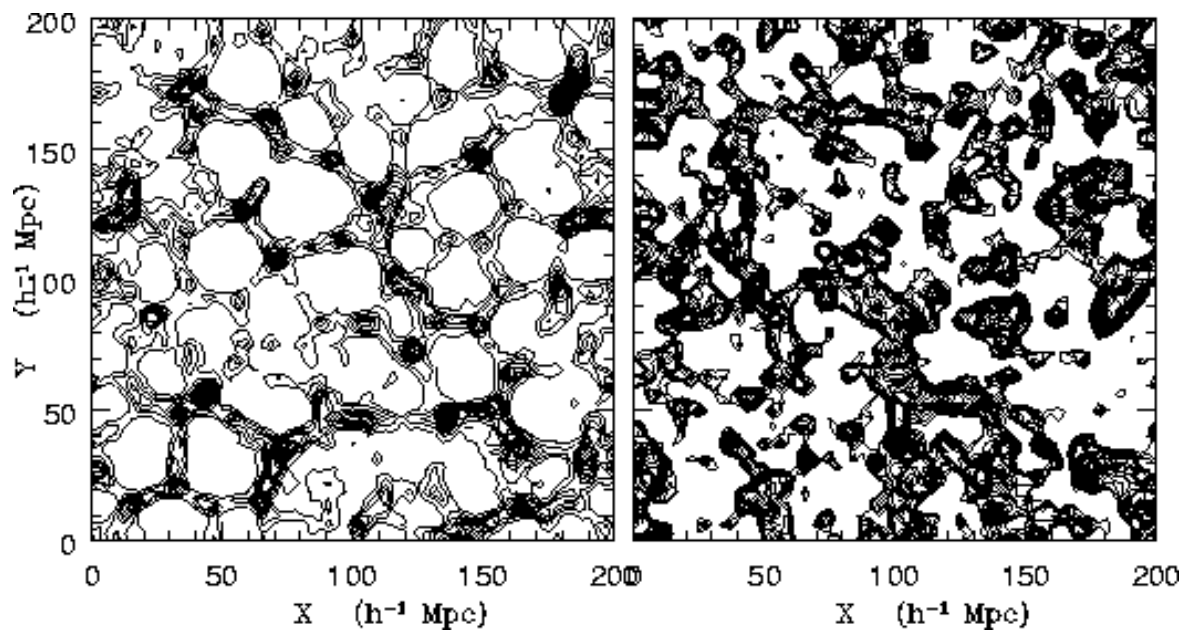}
\caption{A density field with a characteristic cellular geometry (left), 
and its counterpart with the same power spectrum $P(k)$, yet scrambled 
phases. The contour levels of the righthand frame are chosen such that 
only positive $\delta$ levels are indicated. At the lefthand side the 
density contours range from $\delta=0.75$ to $\delta=10.$ in 20 steps. 
Linear contours in both left and right frame. After a suggestion of 
Alex Szalay.}
\end{figure}

\section{The Cosmic Abacus: quantifying structure}
Hence, a substantial amount of observational evidence seems to indicate 
that already at a remarkably young age the Universe shed  
its primordial featureless and pristine complexion, and set out 
to forging the scaffolds for the construction of the patterns that 
pervade our Universe on Megaparsec scales. In order to turn this 
qualitative conclusion to further use, and decide which theoretically 
proposed scenario lay at the basis of the observed patterns, we evidently 
need to find and/or find measures to quantify those aspects of the 
matter and galaxy distribution that are as strongly discriminative 
as possible. Shedding aside the 
highly complex anisotropic patterns, most work has concentrated on the 
first orders of the clustering process, effectively describing the likelihood 
and frequency of over- and underdensities over a range of scales, as well 
as their mutual spatial correlation.

The standard contention is that structure grew from a random 
distribution of density fluctuations $\delta({\bf x})$ whose 
statistical properties are described by a Gaussian distribution function. 
In other words, all its Fourier components ${\delta}({\bf k})$ are mutually 
independent and have a Gaussian distribution, with its average amplitude 
determined by the $\sigma(k)$, usually denoted by the name of ``power 
spectrum'' $P(k)$. Physically, the power spectrum expresses the 
relative average magnitude of the Fourier waves at every relevant scale in 
the constituent spatial density field realization. The relative clout 
of the various waves is of crucial importance determing the outcome and 
character of the final matter organization. A primordial field with 
a blue spectrum with high amplitudes of small scale waves will lead to 
a scenario in which small scale clumps will spring up as first 
discernable objects, while on the other hand a red spectrum will 
yield an evolutionary scenario more resembling the ``top-down'' 
unfolding described earlier. In fact, theoretical work has come up with 
a hoard of analytical power spectra whose shape and amplitude are 
determined by various cosmological factors, global cosmological parameters 
like $\Omega$ and the Hubble parameter $H_o$, but also by the nature of 
the matter, curvature of space, and several other factors. Hence, 
its determination from the observations has such a high priority 
in cosmological research. Hence, the strong incentive towards recovering 
the power spectrum from observations relating 
to the matter distribution over scales ranging from those of the 
large Gigaparsec scales discerned in the microwave background temperature 
fluctuations, through the hundred Megaparsec scales whose fluctuations 
imprints can be measured from the large-scale velocity fields and cluster 
clustering, down to nonlinear scales of a couple of Megaparsec, where it 
is hoped the galaxy distribution still contains sufficient information. 

Determinations of the power spectrum have been preceded by and still go 
hand-in-hand with a huge amount of effort in describing the 
fluctuation field in terms of the spatial Fourier transform of the power 
spectrum, the correlation function $\xi({\bf r})$. In principal, $\xi$ 
contains exactly the same amount of information as $P(k)$, although 
observational errors make it more practical to determine $\xi$ on 
small scales, while its drowning in noise at larger scales make the 
power spectrum the quantity of preference on scales exceeding some 
10 Megaparsec. 

However, once gravitational instability gets hold of the primordial 
field and starts moulding it into density field realizations in which 
higher and higher density peaks collapse to smaller and smaller parts 
of space and low density regions empty themselves while seizing larger 
and larger chunks of the Universe, nonlinear gravitational processes 
start to evoke larger and larger deviations from the initial 
Gaussian distribution function. It therefore becomes more and more 
elaborate to characterize the clustering of matter. Fluctuations over a 
range of scales start 
to influence each other, with for instance small-scale density enhancements 
in large scale overdense regions collapsing earlier than those in more 
barren regions of space. Hence, the various waves start to interact, and 
transfer power between the different scales. Another process contributing 
to power transfer between different scales is the tendency of density 
enhancements and depressions in changing shape, high density regions 
will collapse to more and more anisotropic configurations in the generic 
situation of them not being spherical while low density regions expand 
to a more spherical shape. The ultimate outcome of the gravitational 
evolution is therefore a field that becomes increasingly non-Gaussian, 
in the sense of developing a larger and larger amplitude of higher 
order correlation functions besides the second order correlation 
functions. A lot of work was therefore devoted to determining some 
higher orders of the correlation function hierarchy, but at some 
point this becomes an almost impossible task, the signal being drowned 
in the noise of the observations. Only in the early quasi-linear 
stages of gravitational evolution, when density fluctuations are still 
in the order of $\delta \sim 1$ it is still reasonable to expect 
the lower order correlation functions more or less fully quantifying 
structure. 

However, once gravitational clustering starts to transform the 
density field into one exhibiting a variety of interesting patterns 
over a vast range of scales, once collapsed and virialized density 
clumps start to pop up, any hope of a full statistical quantification 
gets lost. That the power spectrum and correlation functions are not 
fully equiped in quantifying the most conspicuous aspects of the 
emerging large scale structure can be discerned from Figure 4. 
Even only a superficial look at Figure 1 shows how much of an essential 
aspect of the matter distribution is then swept under the carpet. In fact, 
the power spectrum does not contain any information on the foamlike 
morphology of the matter distribution. Figure 4 displays two density 
fields with exactly the same power spectrum. However, while the lefthand 
one exhibits a beautiful foamlike morphology, the righthand one is 
but a featureless Gaussian field. 
A lot of effort has therefore been devoted to developing and 
defining statistical quantities that characterize various aspects 
of the matter distribution, in the hope of them being strongly discriminatory. 
However, a lot of these attempts produce merely heuristic measures, 
which have a poorly understood relationship to the underlying scenario of 
clustering. For instance, minimum spanning trees and percolation 
measures have gone some way into quantifying filamentary structures, 
but as yet their relation to the initial power spectrum of 
fluctations is unclear. Topology measures do have some quantified 
relation to the power spectrum, but it's discriminatory virtues 
and visual clarity are still contentious, while the same can probably be 
stated about Minkowski functionals. Attempts in quantifying 
structure through a hierarchy of fractal dimensions are interesting, 
but useful only over a limited range of scales where the various 
moments of the density field exhibit scaling behaviour. Perhaps one 
of the most interesting and promising approaches, but as yet not fully 
understood, is the description of structure in terms of wavelets. 
Although at first received in the cosmological community with a huge 
grain of scepticism, there are strong indications from other fields of 
physics that they indeed are very suitable characterizations of 
non-trivial patterns in nature (see e.g. Bowman \& Newell 1998). In 
addition though, we should continue to study in more detail the structure 
of the density fields within the well-known realms of Fourier space, 
and assess in a more systematic way the evolution of phases, phase 
distributions and phase correlations through the action of gravity, and 
fill in the meanings behind the buzz-words so often employed. In other 
words, we should start thinking about a holographic analysis of 
the observed as well as simulated matter distributions. 

While cosmologists are pursuing the search for transparent ``measures 
of reality'' (Crosby 1997), a host of information about the 
structure formation scenario casting our Universe can also be 
obtained by a complementary approach, producing theoretical 
realizations of nonlinear matter distributions for a host of 
scenarios and comparing them both by their visual impression as 
well as through the various statistical measures that we can 
find in our toolbox.

\begin{figure}
\centering
\epsfxsize=15.5cm
\epsfbox{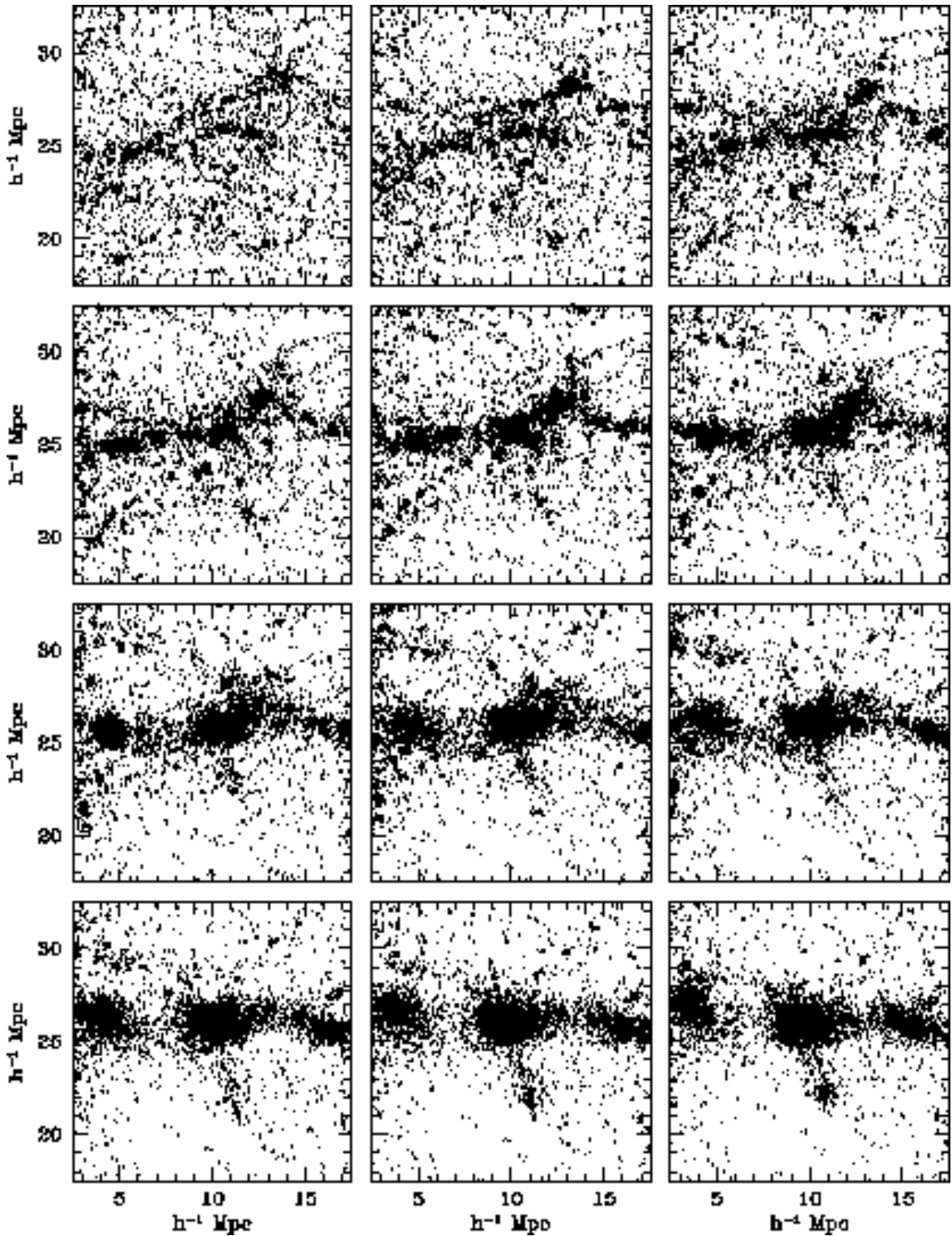}
\caption{An illustration of gravitational clustering and collapse. The 
development of a small region in a $100h^{-1}\hbox{Mpc}$ box, within 
the standard CDM scenario, is followed in a sequence of 12 time steps, 
going from left to right, top to bottom. The final timestep, bottom 
right, should correspond to the present epoch, $a=1$ (From 
Van de Weygaert \& van Albada 1996).}
\end{figure}

\section{Cosmic Pretensions: simulating structure formation}
As it appears to be a forbidding task to infer direct inferences from 
the observed nonlinear galaxy distribution about the valid scenario of 
structure formation, one can also pursue another approach. This 
approach comprises the simulation of nonlinear mass distributions in 
one or more structure formation scenarios. An additional advantage is 
that it not allows a quantitative comparison with reality, through 
a comparison of similar statistical quantifications, but also a 
still very useful qualitative assesment by comparing the visual 
impression of the observed Universe with that in the simulated 
portion of the Universe. In fact, while we have seen that we are 
still failing to characterize striking patterns and properties of 
for instance cellular matter configurations, this is an essential 
tool in the study of large scale structure. 

Progress in these structure evolution simulations have been tremendous 
by sheer of a continous and an acceleraring increase in available computer 
power and available memory space. Several decades ago the early modest 
particle simulations counted at best a few hundred particles, and the 
first influential structure formation simulations of the Cold Dark Matter 
scenario comprised 32,000 particles (Efstathiou et al. 1985), the standard  
simulation at present already contains at least several million particles, 
with state-of-the-art simulations exceeding even a billion particles. 

The basics of structure formation simulations can be shortly summarized. 
A realization of a primordial density and velocity fluctuation field for 
a specific scenario is generated. Its structure is subsequently 
discretized by a finite, yet very large, number of particles, usually 
of equal mass. Evolving this distribution slightly through an analytical 
approximation of matter displacements in the early quasi-linear stage 
of clustering, the Zel'dovich approximation, the stage is set for an 
elaborate sequel through an N-body code that is capable of following 
the full nonlinear evolution by solving for each individual particle 
the equations of motion at a sequence of timesteps. Oversimplifying 
various different ways of performing these N-body simulations, the 
usual strategy is to interpolate the particle distribution to yield 
a density field which in turn yields the underlying gravitational 
potential field through solving the Poisson equation. After having 
done so, we can interpolate back to the particles to yield their 
gravitational acceleration, and subsequently their velocity. By 
doing this at a myriad of timesteps, for a huge number of particles, 
we hope to obtain an idea of the intricacies of nonlinear 
gravitational clustering. An illustration of such a sequence of 12 
timesteps is given in Figure 5, showing the evolution of a clump 
of matter with a mass in the range of that of groups of galaxies 
like our own Local Group. The continuous accretion and concentration 
of ever larger amounts of matter is clearly born out, as is the 
rather anisotropic nature of the whole process. 

\begin{figure}
\centering
\epsfxsize=15.5cm
\epsfbox{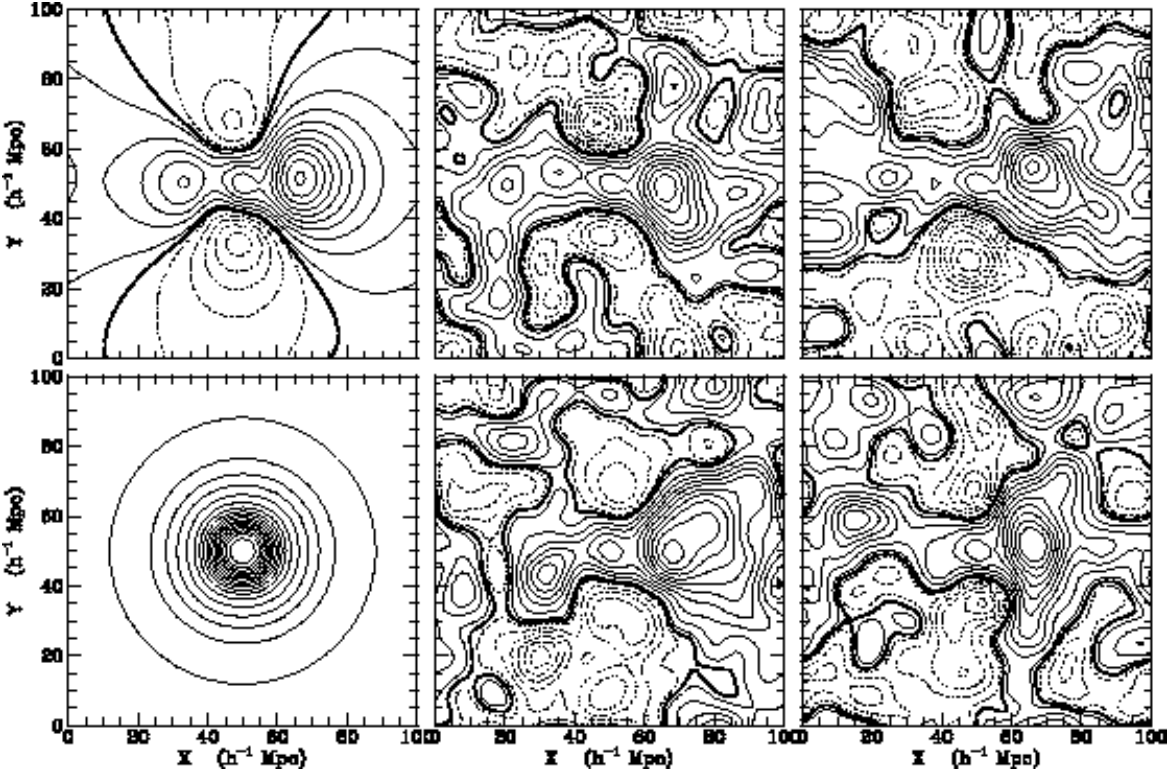}
\caption{The variance of constrained random field realizations illustrated 
by means of contour density maps in a $5h^{-1}\,\hbox{Mpc}$ thick central 
slice in a $100h^{-1}\,\hbox{Mpc}$ box. See text (from Van de Weygaert 
\& Bertschinger 1996).}
\end{figure}

A lot of effort is devoted to simulate the Universe to great detail 
over a range as large as possible, in an attempt to adhere to the 
purest definition of ``simulation'', trying to reproduce reality in 
all its aspects (see e.g. the work of the Virgo consortium, e.g. 
Pearce \& Couchman 1997). While this obviously is necessary if ever we wish to 
have confidence in our model Universes describing the real Universe, 
it automatically biases scientific efforts towards trying to 
concentrate huge amounts of manpower and financial resources 
towards that one goal. However, there is still ample space for a 
complementary approach, arguably as necessary towards obtaining an 
understanding of the action 
of gravity in shaping our cosmic environment. 

While the massive state-of-the-art simulations try to reproduce 
the real Universe, and therefore comprise all the different detailed 
processes whose interactions conspire to produce the final mass 
distribution, a full understanding cannot be reached without trying 
to isolate and understand the various relevant processes and 
physical factors. In other words, is it possible to pursue a more 
laboratory oriented approach, not trying to reproduce the Universe 
in all its charms, but rather concentrating of one or a few 
supposedly relevant factors. The insight provided through such 
an approach will reveal more clearly the relative importance of 
physical processes, quantities and matter configurations in the 
full-blown large simulations. 

\begin{figure}
\centering
\epsfxsize=15.5cm
\epsfbox{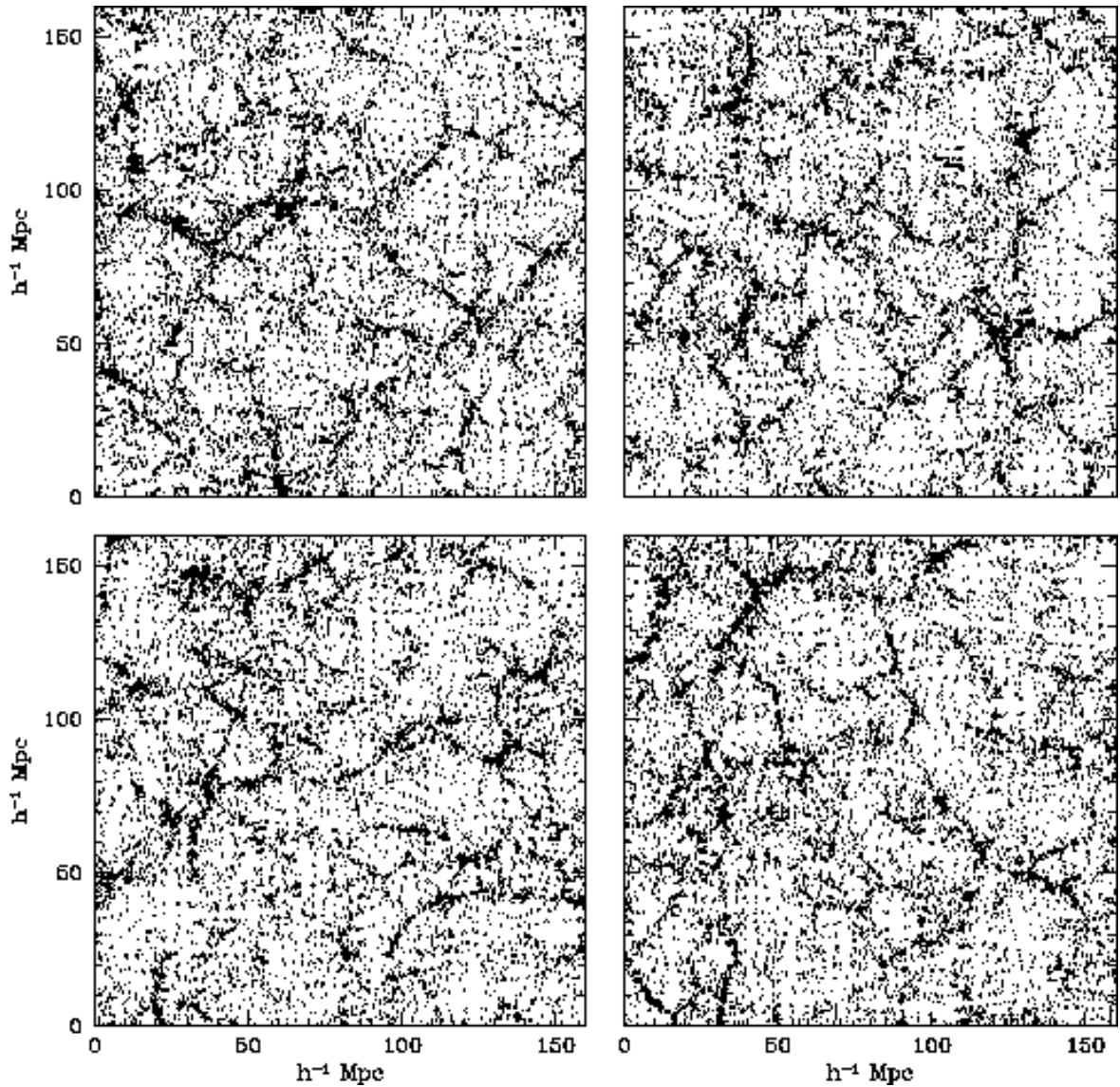}
\caption{Four constrained realizations of the nonlinear evolution of 
our local Universe (see text, Van de Weygaert \& Hoffman 1998). Based 
on a Wiener reconstruction of the local density field on the basis 
of the Mark III catalogue of galaxy peculiar velocities (Willick et al. 
1997). The extended concentrations of mass discernable towards 
the upper lefthand quarter of all four realizations may be regarded 
as the kins of the Great Attractor, while to the lower righthand side 
something akin to the Perseus-Pisces supercluster can be seen.}
\end{figure}

The laboratory approach hinted at above is what we strive for through 
simulations based on constrained field realizations (see Bertschinger 1987, 
Hoffman \& Ribak 1991, Van de Weygaert \& Bertschinger 1996). Figure 6 
illustrates the basic ideas behind such constrained field realizations. 
Its top lefthand panel 
illustrates the imposed set of constraints by means of the {\it mean field}  
${\bar f}$, which is set by this particular set of constraints and constraint 
values (see Van de Weygaert \& Bertschinger 1996). The constraint set 
involves the peculiar velocity and tidal field at the central location 
of the simulation box. To yield genuine realizations of a random field 
obeying the constraint set, a {\it residual field} is added to the mean 
field. This residual field contains the fluctuations intrinsic to a field 
with the relevant power spectrum. 
The four central and righthand panels display four different 
random realizations adhering to these constraints. While the overall 
pattern of the mean field can clearly be recognized in all four random 
realizations, they also show where and on what scale the realizations 
can differ from one to the other. Clearly, around the location at which 
the constraints have been specified, the variation amongst the realizations 
is negligible. Further out it approaches the generic variation expected 
for unconstrained random fields. Also, smaller scales are far less affected 
than the scale at which the constraints is set, in this case a Gaussian 
scale of $R_f = 5h^{-1}\,\hbox{Mpc}$. To quantify and summurize the 
variations between the different realizations, the bottom lefthand panel 
is the contourmap of the value of the variance of the field realizations 
inside the slice, running from 0.0 at the centre to $\sigma_0\approx 0.95$ 
at the edge of the box.
Note that in this particular case, with the constraint concerning the 
value and configuration of the tidal field at the centre of the box, 
the presence of a strong straining tidal component along the $x$-axis, 
in combination with compensating compressing components along the 
$y$- and $z$-axis automatically implies a pronounced quadrupolar mass 
distribution, ultimately evolving into a configuration of two massive 
clumps connected by an elongated thinner bridge in between them. Hence, 
this explains the frequently mentioned and observed connection between 
clusters of galaxies and filaments (see e.g. Bond, Kofman \& Pogosyan 
1996) !

A particular insightful application of this idea is by taking the 
constraints from observed reality. For example, the density field 
in the local Universe as implicated by the local cosmic 
flow field. This local flow field can for instance be obtained 
through interpolation of the measured galaxy peculiar velocities listed 
in the Mark III catalogue (e.g. Willick et al. 1997). Arguably the best 
estimate of the corresponding linear density field in the local Universe, 
the one with the highest signal-to-noise level, is obtained through the 
application of the Wiener filter technique developed by Hoffman, Zaroubi 
and collaborators (see Zaroubi et al. 1995). The latter is the de facto 
mean field of all resulting realizations implied by the measured cosmic 
flow field and corresponding measurement errors. Subsequently, on the premise 
of a CDM Universe with $\Omega=1.$, four different realizations were  
generated by adding appropriately constrained noise fields to the 
Wiener filter reconstructed field. To this end we invoked the Hoffman-Ribak 
constrained random field recipy (Hoffman \& Ribak 1991, van de Weygaert 
\& Bertschinger 1996). The resulting linearly extrapolated density 
fields are used as initial density field realizations, and their 
further nonlinear evolution is followed by means of a P$^3$M N-body 
code. The outcome at an expansion factor $a=0.8$ for the four 
different realizations are displayed in the four panels, each panel 
representing the particle distribution in a $10h^{-1}\,\hbox{Mpc}$ 
thick slice through the centre of the box, which corresponds to 
our position in the cosmos. The extended mass concentration to the 
lefthand side should correspond to possible realizations of the 
Great Attractor region, while towards the lower righthand side one 
can recognize a matter concentration that in the case of our real 
Universe is called Perseus-Pisces supercluster (see Van de Weygaert \& 
Hoffman 1998).

Clearly, in this way we will be able to systematically explore 
various physical effects at work in our local neighbourhood. Knowledge 
obtained from these assessments can then be incorporated in an 
analysis of a full-blown superduper simulation. In this way we hope 
to crawl further and further towards a configuration resembling 
as closely as possible the large-scale environment in which we 
live, setting the scene for solving maybe the most mesmerizing 
riddle of them all, the Holy Grail of 20th century cosmogony, 
the creation, the rise, evolution and growth of those jewels 
in the ``crown of creation'' (Jefferson Airplane 1968), the {\it galaxies}. 

\begin{acknow}
I would like to thank Bernard Jones for useful suggestions and friendly 
comments, and Yehuda Hoffman for his permission to include figure 6 
in advance of publication. In addition, I am grateful to the 
editors for their almost infinite patience, and to Vincent Icke 
for the Measure of Reality.
\end{acknow}

\end{document}